\documentclass[%
reprint,
superscriptaddress,
amsmath,amssymb,
aps,
prb,
noeprint,
floatfix,
]{revtex4-2}

\usepackage[pdftex]{graphicx}
\usepackage{bm}
\usepackage{amsmath}
\usepackage{textcomp}
\usepackage{xr-hyper}
\usepackage{hyperref}
\usepackage[all]{hypcap}
\usepackage{xcolor}
\usepackage{placeins}
\usepackage{enumitem}
\allowdisplaybreaks

\makeatletter
\newcommand*{\balancecolsandclearpage}{
  \close@column@grid
  \cleardoublepage
  \twocolumngrid
}
\makeatother

\begin{document}

\title{Impact of current-induced magnons on spin-orbit torque analysis}
\makeatletter
\let\newtitle\@title
\let\newauthor\@author
\let\newdate\@date
\makeatother

\author{Tam\'as Prok}
\affiliation{Department of Physics, Institute of Physics, Budapest University of Technology and Economics, M\H{u}egyetem rkp.\ 3., H-1111 Budapest, Hungary}
\affiliation{MTA-BME Correlated van der Waals Structures Momentum Research Group, M\H{u}egyetem rkp.\ 3., H-1111 Budapest, Hungary}
\affiliation{Zernike Institute for Advanced Materials, University of Groningen, Groningen, The Netherlands}

\author{Jan Hidding}
\affiliation{Zernike Institute for Advanced Materials, University of Groningen, Groningen, The Netherlands}

\author{Szabolcs Csonka}
\affiliation{Department of Physics, Institute of Physics, Budapest University of Technology and Economics, M\H{u}egyetem rkp.\ 3., H-1111 Budapest, Hungary}
\affiliation{MTA-BME Superconducting Nanoelectronics Momentum Research Group, M\H{u}egyetem rkp.\ 3., H-1111 Budapest, Hungary}
\affiliation{HUN-REN Centre for Energy Research, Institute of Technical Physics and Materials Science, Konkoly Thege Miklós út 29-33, 1121 Budapest, Hungary}

\author{P\'eter Makk}
\affiliation{Department of Physics, Institute of Physics, Budapest University of Technology and Economics, M\H{u}egyetem rkp.\ 3., H-1111 Budapest, Hungary}
\affiliation{MTA-BME Correlated van der Waals Structures Momentum Research Group, M\H{u}egyetem rkp.\ 3., H-1111 Budapest, Hungary}

\author{Marcos H.\ D.\ Guimar\~aes}
\affiliation{Zernike Institute for Advanced Materials, University of Groningen, Groningen, The Netherlands}

\author{Endre T\'ov\'ari}
\email{tovari.endre@ttk.bme.hu}
\affiliation{Department of Physics, Institute of Physics, Budapest University of Technology and Economics, M\H{u}egyetem rkp.\ 3., H-1111 Budapest, Hungary}
\affiliation{MTA-BME Correlated van der Waals Structures Momentum Research Group, M\H{u}egyetem rkp.\ 3., H-1111 Budapest, Hungary}

\date{\today}
\begin{abstract}
The second-harmonic Hall technique is a widely used, sensitive method for studying the spin-orbit torques generated by charge current. It exploits the dependence of the Hall resistance on the magnetization direction, although thermal phenomena also contribute. Historically, deviations from the expected magnetic field dependence have usually been neglected. Based on our studies on permalloy/platinum bilayers, we show that a counterpart to the magnon-related spin-flip unidirectional magnetoresistance - known to appear in the second-harmonic longitudinal resistance - appears in the Hall data, and that describing the results in a wide field range with these contributions is essential to accurately estimate the torques. 
\end{abstract}

\maketitle
\counterwithin*{equation}{part}
\stepcounter{part}
\renewcommand{\theequation}{\arabic{equation}}

\section{Introduction}\label{sec:intro}

Two key pathways for advancing magnetic random-access memory are spin transfer torque (STT) and spin-orbit torque (SOT) \cite{Dieny2020, Krizakova2022}. In STT a spin current is created by driving a charge current through a ferromagnetic (FM) layer, which flips the state of the FM layer that constitutes the memory bit \cite{Bhatti2017, Liu2019}. Spin currents can also be generated in a non-magnetic material with strong spin-orbit coupling, via charge-to-spin conversion mechanisms such as the Rashba-Edelstein (also called the inverse spin galvanic) and the spin Hall effects \cite{Sinova2015}. The resulting spin-orbit torque can also be used to switch the magnetization of an adjacent FM \cite{Bi2019, Manchon2019, Nguyen2024}. 

For the analysis of torques, spin-torque ferromagnetic resonance \cite{Liu2011,Karimeddiny2020} and second-harmonic Hall (2HH) measurements \cite{Garello2013, Hayashi2014} are often used \cite{Fan2013,Dyakonov2017}, from heavy metals \cite{Mosendz2010, Pi2010, Demasius2016, Wang2022} to van der Waals crystals \cite{Yang2022, Guillet2024, Pandey2024} or their combinations \cite{Guimarães2018, Shao2016, Macneill2017b, Alghamdi2019, Hidding2020, Bainsla2024}. The 2HH technique is sensitive to spin-orbit fields generated by an AC current applied to the structure. As these fields slightly alter the magnetization direction and thus the magnetoresistance, voltage components appear that are quadratic with the current $I$, and thus can be detected at double frequency, $2\omega$ \cite{Garello2013, Hayashi2014, Avci2014, Shao2016}. 

Several other phenomena leading to a double-frequency voltage signal complicate the analysis of 2HH measurements. Various thermal effects from Joule heating $\propto I^2$ such as the anomalous Nernst effect are prominent among them \cite{Bauer2012, Zink2022}. However, these are sometimes inadequate to fully describe the $2\omega$ signals, which is generally circumvented by fitting to only part of the data. Here we go beyond this approach: we discuss modifications to the original model and show that the unusual behavior of the Hall data is reminiscent of the magnon-related spin-flip unidirectional magnetoresistance (UMR) that appears in the longitudinal data \cite{Avci2015, Langenfeld2016, Li2017, Avci2018, Kodama2024}. We compare two approaches to describe our observations on permalloy/platinum (Py/Pt) bilayers, and for the second approach, demonstrate satisfactory fits to the 2$\omega$ measurements over a broad range of magnetic field values. Our analysis leads to significant quantitative changes in the extracted SOT paramaters, and highlights the importance of taking UMR-like contributions into account in future studies.

\section{Experimental results}\label{sec:results}

We have carried out measurements on eight ferromagnet/normal metal Hall bar devices at room temperature, with qualitatively similar results. The devices were made of 6~nm permalloy (Py) capped by 3~nm platinum (Pt), and contacted with Ti/Au electrodes on a SiO$_2$-covered wafer. The measurements presented in the main text were taken on the sample shown in Fig.~\ref{fig1}a); data on other samples can be found in the Supplementary Information (SI \cite{Comment2}) Section~\ref{subsec:extras}. We applied $2.5\,\mathrm{m A}$ rms-amplitude AC current along the length of the Hall bar, and measured the longitudinal $V_{xx}$ and transverse (Hall) $V_{xy}$ voltages by low-frequency lock-in technique as shown in Fig.~\ref{fig1}a). Further measurement parameters and sample fabrication are detailed in Section~\ref{sec:methods}. We studied the first and second-harmonic ($2\omega$) longitudinal and Hall resistances $R_{xx,xy}$, $R_{xx,xy}^{2\omega}$ as a function of out-of-plane magnetic field as well as in-plane rotation in a constant magnetic field. We denote the polar angles of the dimensionless magnetization $\bm{M}$ and the magnetic field $\bm{H}$ relative to the out-of-plane direction ($z$) by $\theta$ and $\theta_H$, respectively, and the azimuthal angles relative to the current direction $x$ by $\varphi$ and $\varphi_H$ as shown in Figs.~\ref{fig1}a,b). Since $\mathrm{Py}$ is an easy-plane ferromagnet, $\varphi\approx \varphi_H$ for all values of magnetic field used in this work ($\mu_0 H > 15~\mathrm{mT}$). 

We plot measurements of $R_{xy}$ as a function of out-of-plane field $H_z$ in Fig.~\ref{fig1}c). Since this is the hard axis, the hysteresis was negligible. We observe a clear anomalous Hall effect (AHE) with saturation at $\mu_0 |H_z| > 0.7~\mathrm{T}$, which reflects the behavior of the $z$-component of the magnetization. $M_z$ becomes finite as $\bm{M}$ is gradually tilted out of the easy ($xy$) plane with increasing $|H_z|$, then becomes parallel with $\bm{H}$. The shallow linear background originates in the ordinary Hall effect (OHE). The resistance offset and the parts that are even as a function of $H$ (such as the dip at $H=0$) are attributed to the mixing of $R_{xx}$. 

In Fig.~\ref{fig1}d) we show $R_{xx}$, $R_{xy}$ as a function of $\varphi_H$ taken by rotating the sample in constant in-plane field. The $\pi$-periodic oscillations originate from the anisotropic magnetoresistance (AMR) and its sister phenomenon, the planar Hall effect (PHE) \cite{Ritzinger2023}. We summarize these effects in the Hall and longitudinal resistances at a given field, neglecting offsets and the OHE, as
\begin{subequations}
	\label{eq:R} 
	\begin{eqnarray}
    R_{xy} &=& R_A \cos \theta + R_P \sin^2 \theta \sin 2 \varphi
	\label{eq:Ra}
	\\
	R_{xx} &=& R_0 + R_a \sin^2 \theta \cos^2 \varphi. \label{eq:Rb}
	\end{eqnarray}
\end{subequations}
In $R_{xy}$ the first term is the AHE, and its magnitude is illustrated in Fig.~\ref{fig1}c). We correct for the small contribution of the OHE by extrapolating linear fits from the saturation region (nearly horizontal red lines) to zero field (blue line). The second terms in Eqs.~\ref{eq:Ra}, \ref{eq:Rb} describe PHE and AMR. The corresponding fits can be found in Fig.~\ref{fig1}d) in red and blue. Since the aspect ratio here is 1, their amplitudes are expected to be the same, $R_P = R_a/2$ \cite{Ritzinger2023}. We observe a $\sim 10\%$ deviation from this relation which may be due to an imperfect geometry or inhomogeneities in the sample.

\begin{figure}[!t]
\begin{center}
	\includegraphics[page=1]{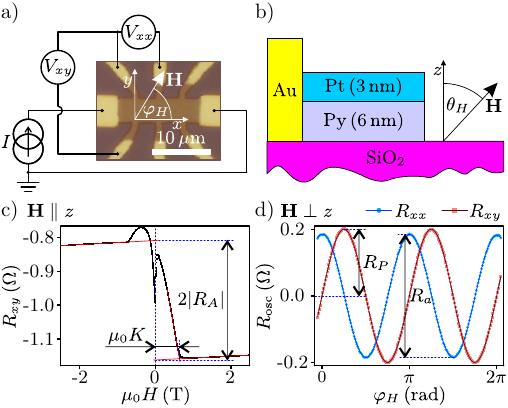}
	\caption{
	\textbf{Measurement geometry and first harmonic data.}
	a) Optical image of the device and diagram of the measurement setup, also showing the in-plane magnetic field direction.
    b) Illustration of the device composition and the polar angle of the field $\bm{H}$.
    c) Hall resistance $R_{xy}$ as a function of out-of-plane field ($\theta_H = 0$). The red lines are fits to the saturation regions and to a part of the transition region. The anomalous Hall resistance ($R_A \approx -179\,\mathrm{m \Omega}$) and the anisotropy field ($\mu_0 K \approx 0.685\,\mathrm{T}$) based on the lines' crossing points are highlighted. d) The oscillating parts of $R_{xx}$, $R_{xy}$ (blue and red symbols, respectively) for in-plane rotation ($\theta_H = \pi/2$) at $\mu_0 H = 0.186\,\mathrm{T}$. Solid lines are fits to Eq.~\ref{eq:R} after including small horizontal offsets. The constant backgrounds that were removed are $81.8$ and $-0.98\,\mathrm{\Omega}$, respectively.}
	\label{fig1}
\end{center}
\end{figure}

Additional magnetoresistance (MR) effects may contribute to the terms labelled as AHE, PHE and AMR above, such as spin Hall MR \cite{Kim2016}, Hanle MR \cite{Velez2016}, anomalous Hall MR \cite{Yang2018}, and the geometrical size effect \cite{Rijks1997, Kobs2011, Kang2020}. The isotropic offset $R_0$ is affected by magnon MR (MMR) \cite{Mihai2008, Kobs2011}. However, they do not change the $(\theta,\varphi)$-dependence in Eq.~\ref{eq:Ra} and the $\varphi$-dependence for in-plane rotation ($\theta = \pi/2$) in Eq.~\ref{eq:Rb} which are later exploited for determining SOT, and therefore can be collectively described by $R_P$ and $R_a$. $\varphi \approx \varphi_H$ applies well for the studied fields which is confirmed by the goodness of the fits to Eq.~\ref{eq:R} in Fig.~\ref{fig1}d), although there is a weak in-plane easy-axis magnetic anisotropy in our devices (see SI Sec.~\ref{subsec:SW2} and also reference \cite{Macneill2017a}).

\begin{figure*}[!tbh]
\begin{center}
	\includegraphics[page=2]{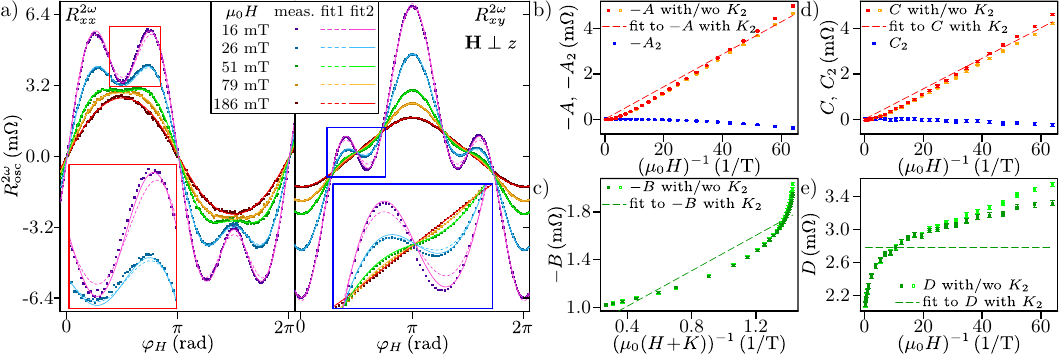}
	\caption{
    \textbf{Second harmonic Hall measurements.} 
	a) $R_{xx}^{2\omega}$, $R_{xy}^{2\omega}$ with in-plane field rotation ($\theta=\theta_H=\pi/2$) for selected values of the absolute field $\mu_0 H$ after averaging clockwise and counter-clockwise rotations and removal of constant offsets. The dashed lines (fit1) are fits according to Eq.~\ref{eq:R2w} and $\varphi=\varphi_H$. The solid lines (fit2) include the anisotropy $K_2$ as corrections to $\varphi$ (Eq.~\ref{eq:SW5}) and include the terms of Eq.~\ref{eq:R2w_X2}. Insets are zooms of the highlighted rectangles. b-e) The coefficients $-A,-A_2,C,C_2$ and $-B,D$ as a function of $(\mu_0 H)^{-1}$ and $(\mu_0(H + K))^{-1}$, respectively.  The dashed lines are linear fits according to Eq.~\ref{eq:R2wComps}.
	}\label{fig2}\end{center}
\end{figure*}

In the following we discuss low-frequency 2$\omega$ measurements for quantifying SOT. The spin Hall effect in Pt produces spins with orientation $\bm{\sigma}$ in the $-y$ direction at the Py/Pt interface. They, along with the external $\bm{H}$, exert torques on the magnetization $\bm{M}$. On the time scale of the lock-in measurement, the precession of $\bm{M}$ is completely damped, and the torques can be described by effective fields. They slightly tilt the magnetization out of the equilibrium direction (see Eqs.~\ref{eq:SW2}, \ref{eq:SW3} in the SI) with the frequency $\omega$ of the current. The modulation $\Delta \theta , \Delta \varphi$ of the angles in Eq.~\ref{eq:R} thus leads to a $2\omega$ out-of-phase component in the voltage signals \cite{Hayashi2014}. In this high-symmetry system the second-harmonic resistances measured as a function of $\varphi_H$ for $\theta_H=\pi/2$ and plotted in Fig.~\ref{fig2}a) can be fitted according to 
\begin{subequations}
	\label{eq:R2w} 
	\begin{eqnarray}
    \label{eq:R2wa}
	R_{xy}^{2\omega}(\varphi) &=& A \cos 2\varphi \cos \varphi  + B \cos \varphi + \mathrm{const}
	\\
    \label{eq:R2wb}
	R_{xx}^{2\omega}(\varphi) &=& C \sin 2\varphi \cos \varphi + D \sin \varphi + \mathrm{const.}
	\end{eqnarray}
\end{subequations}
As detailed in SI Section~\ref{subsec:SOT} and Refs.~\citenum{Garello2013, Avci2014, Shao2016}, the terms with amplitudes $A,C$ are attributed to the sum of the effective magnetic field $H_\mathrm{FL}$ of the field-like SOT and the Oersted field $H_\mathrm{Oe} = I_\mathrm{Pt}/(2W)$ generated by the current in Pt ($W$ is the width of the Hall bar) \cite{Macneill2017a}. $H_\mathrm{FL} $ and $ H_\mathrm{Oe}$ always appear together and cannot be easily distinguished. For simplicity, we will write $H_\mathrm{FL}$ instead of $H_\mathrm{FL} + H_\mathrm{Oe}$. Coefficient $B$ originates from the anti-damping SOT field $H_\mathrm{AD}$ and the anomalous Nernst effect $R_\mathrm{ANE}$. The ANE Hall voltage is related to the Joule-heating-induced thermal gradient $\bm{\nabla} T$ in the sample by $V_\mathrm{ANE,y} \propto (\bm{\nabla} T \times \bm{M})_y$ where $\bm{\nabla} T \propto I^2(t)$ has $2\omega$ frequency and an out-of-plane component. $D$ is attributed to ANE through $(\bm{\nabla} T \times \bm{M})_x$. Finally, constant terms (independent of $\varphi$ and $H$) are possible due to in-plane components of $\bm{\nabla} T$ via the Seebeck effect. Other thermal effects with different $\varphi$-dependence are possible \cite{Avci2015, Ki1966, Jayathilaka2015, Reimer2017}, but these are negligible in our data. The coefficients $A$--$D$ depend on the field as follows (Refs.~\citenum{Garello2013, Avci2014, Shao2016} and SI Secs.~\ref{subsec:SW}, \ref{subsec:SOT}):
\begin{subequations}
	\label{eq:R2wComps} 
	\begin{eqnarray}
	A(H) &=& R_P \frac{H_\mathrm{FL} }{H},
	\label{eq:R2wCompsa}
	\\
	B(H) &=& \frac{R_A}{2} \frac{H_\mathrm{AD}}{H + K} 
	+ R_\mathrm{ANE},
    \label{eq:R2wCompsb}
    \\
    C(H) &=&  \frac{-R_a}{2} \frac{H_\mathrm{FL} }{H},
    \label{eq:R2wCompsc}
    \\
    D(H) &=& R_\mathrm{ANE2}.
    \label{eq:R2wCompsd}
	\end{eqnarray}
\end{subequations}
We will discuss further contributions, some of which can be built into the above terms, further below and in SI Sec.~\ref{subsec:possible} (see also references \cite{Uchida2010, Slachter2010,Pu2006} therein).

The $\varphi$-dependence of $R_{xx}^{2\omega}$, $R_{xy}^{2\omega}$ and fits corresponding to Eq.~\ref{eq:R2w} are shown in Fig.~\ref{fig2}a) at selected fields. Most rotations were performed between 15-100~mT, since the effect of SOT is strongest at low $H$. Just as when fitting AMR and PHE, we allowed for an offset in $\varphi$ (on the order of 5°) to account for the misalignment of the sample and the magnet. The field-dependence of the amplitudes $A$-$D$ is shown in Fig.~\ref{fig2}b-e) by orange and light green markers. According to Eq.~\ref{eq:R2wComps}, coefficient $D$ is expected to be constant, while $A,C$ and $B$ are linear as a function of $(\mu_0 H)^{-1}$ and $(\mu_0(H + K))^{-1}$, respectively. However, the observations do not match Eq.~\ref{eq:R2wComps} well. For $A$ and $C$, the measured data departs from a linear behavior for low $1/H$ (high fields), while for $B$ the trend is the opposite and the data is nonlinear for high $(H + K)^{-1}$ (low fields). Therefore, it is unlikely that both deviations come from demagnetization and domain wall formation at low $H$. This conclusion is supported by the lack of sudden jumps from domain switching, the overlap of the clockwise and counterclockwise (decreasing/increasing $\varphi_H$) rotations, and the goodness of the fits to Eq.~\ref{eq:R2w} in Fig.~\ref{fig2}a) (dashed lines). The difference between the latter and the data is relatively small even at 16~mT, and the error range of $A,B,C,D$ (see panels (b-e)) are comparable to the symbol size. We have consistently observed such character in all our samples, as illustrated in SI Sec.~\ref{subsec:extras}. Therefore, Eq.~\ref{eq:R2wComps} does not describe our observations adequately. The systematic deviations give us an opportunity to reveal contributions from other, mainly magnon-related processes.

\section{Discussion}

To account for the mismatch between the data and the fitting, we have examined several extensions of the basic model. For example, we have considered easy axis shape anisotropy for the Hall bar. As detailed in SI Section~\ref{subsec:SW2}, from the first harmonic resistances we estimate its strength $\mu_0 K_2$ to be on the order of 1 mT. It also introduces new oscillatory terms with amplitudes $A_2, C_2$ and improves the fits to the second harmonics as shown by the solid lines (labeled fit2) in Fig.~\ref{fig2}a). However, $A,B,C,D$ are weakly affected as shown in red and dark green markers in Fig.~\ref{fig2}b-e), and the differences to the field-dependence governed by Eq.~\ref{eq:R2w} (dashed lines) remain present. Several other effects may appear in metallic bilayers, but those corrections were also deemed insignificant, as discussed in SI Sec.~\ref{subsec:possible}.  

We attribute the deviations in $C$ and $D$ to unidirectional magnetoresistance (UMR) \cite{Avci2015, Langenfeld2016, Li2017, Avci2018, Kodama2024}, and we propose that a counterpart of similar origin also appears in $R_{xy}^{2\omega}$ measurements, in coefficients $A$ and $B$. UMR can originate from \textit{(i)} the dependence of the normal metal / FM interface resistance on the direction of $\bm{\sigma}$ relative to $\bm{M}$, \textit{(ii)} the spin-dependence of the bulk conductivity of the ferromagnet, and \textit{(iii)} from magnon-related effects \cite{Avci2018}. Its contribution to the longitudinal resistance $R_{xx}^{2\omega}$ is proportional to $\bm{\sigma} \cdot \bm{M} =- \sin \theta \sin \varphi$ \cite{Avci2015}. The first two \textit{(i,ii)} are both independent of $H$ and are together called the spin-dependent (SD) UMR. In contrast, the final \textit{(iii)}, so-called spin-flip (SF) UMR depends on $H$. It is attributed to the generation and annihilation of magnons through angular momentum transfer from the spins generated by the current in the normal metal \cite{Langenfeld2016, Li2017, Avci2018, Kodama2024}, which is also supported by its correlation with $H_\mathrm{AD}$ \cite{Avci2015b}. Through the modulation of electron-magnon scattering (MMR) or potentially of the length of the magnetization\cite{Noel2024a, Noel2024b}, this leads to a resistance contribution that depends on the current, and on the magnetization via $\bm{\sigma} \cdot \bm{M}$. It decays with increasing field since $H$ opens a gap in the magnon spectrum. Besides the term proportional to $\sin \varphi$, higher frequencies with $\varphi$ have been observed \cite{Avci2018, Xu2019}. Surprisingly, a corresponding correction to the Hall signal has not been reported in diffusive systems to our knowledge, only in bilayers based on topological insulators \cite{Yasuda2016, Yasuda2017}. In the following we present two interpretations on why UMR can appear in 2HH measurements.

\textit{(1) The first approach: symmetry considerations.} In AMR the resistance depends on the direction of magnetization. Its two-fold rotational symmetry leads to its transverse counterpart, PHE \cite{Ritzinger2023}. In contrast, MMR is isotropic and leads to SF-UMR through the modulation of the magnon population $\propto \sin \varphi$, and a transverse part is not expected. However, higher frequencies in $\varphi$ also exist and have been attributed to the combination of SF scattering and AMR\cite{Xu2019}, which implies a similar effect through PHE. Therefore, we propose that $R_{xy}^{2\omega}$ can have components due to the effect which produces parts of SF-UMR in $R_{xx}^{2\omega}$. Including the phase-shift in $\varphi$ (like between AMR and PHE) and accounting for higher frequency terms with $\varphi$, all four coefficients in Eq.~\ref{eq:R2wComps} are to be extended:
\begin{subequations}
	\label{eq:R2wComps2} 
	\begin{eqnarray}
	A(H) &=& R_P \frac{H_\mathrm{FL} }{H} + A_u (H), 
    \label{eq:R2wComps2a}
	\\
	B(H) &=& \frac{R_A}{2} \frac{H_\mathrm{AD}}{H + K} + B_u (H) + R_\mathrm{ANE}.\label{eq:R2wComps2b}
    \\
    C(H) &=&  \frac{-R_a}{2} \frac{H_\mathrm{FL} }{H} + C_u (H) ,
    \label{eq:R2wComps2c}
    \\
    D(H) &=& D_u (H) + R_\mathrm{ANE2}.
    \label{eq:R2wComps2d}
	\end{eqnarray}
\end{subequations} 
The terms $A_u - D_u$ are the field-dependent SF-UMR contributions and their Hall counterparts. The specific form of their decay has been proposed to be of the form $H^{-p}$ \cite{Avci2018, Xu2019} or $(H+H_0)^{-1}$ \cite{Xu2019}, and was even calculated using Boltzmann's formalism in topological insulators \cite{Yasuda2016, Yasuda2017} exploiting spin-momentum locking. In this (first) approach we use 
 \begin{equation}
 \label{eq:R2w_Xu0}
     X_u (H) = X_{u0} \left( \frac{K}{H} \right) ^p
 \end{equation}
where $X=A,B,C,D$ \cite{Comment1}, since this $H$-dependence describes our data considerably better than $(H+H_0)^{-1}$. The field-independent contributions of SD-UMR and its counterpart to $B,D$ are now included in $R_\mathrm{ANE}, R_\mathrm{ANE2}$. 

\begin{figure}[!t]
\begin{center}
	\includegraphics[page=3]{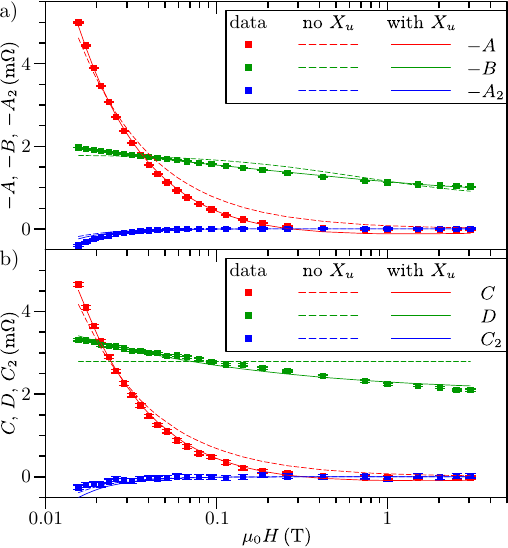}
	\caption{
    \textbf{Comparison of fits without and with $H^{-p}$-type $X_u$ corrections.}
    a) The amplitudes of $R_{xy}^{2\omega}$ and b) $R_{xx}^{2\omega}$ plotted as square symbols, determined from $\varphi_H$-rotations using Eqs.~\ref{eq:R2w}, \ref{eq:R2w_X2}. Fits to the field-dependence without $X_u$ are dashed lines (Eqs.~\ref{eq:R2wComps}, \ref{eq:R2w_X2b}), fits with $X_u$ following Eqs.~\ref{eq:R2wComps2}, \ref{eq:R2w_X2b}, \ref{eq:R2w_Xu0} are solid lines. The field axis is logarithmic for visibility. 
	}\label{Fig3}\end{center}
\end{figure}

In Fig.~\ref{Fig3} we show the results of the fits to the experimentally determined values of $A$--$D$ using Eqs.~\ref{eq:R2wComps2} and ~\ref{eq:R2w_Xu0} as solid lines. They are a significant improvement over the dashed lines that represent the fits lacking $X_u$ (Eq.~\ref{eq:R2wComps}), especially in amplitude $D(H)$, and the residuals are at least an order of magnitude smaller for most fields. However, at high fields $A,C$ do not converge to zero; in fact, the UMR contribution has been observed to deviate from Eq.~\ref{eq:R2w_Xu0} \cite{Chen2022}, but including $H$-linear terms as mentioned therein does not improve the results.

The fit parameters are summarized in Table~\ref{tbl:1}. Firstly, $H_\mathrm{FL}$ is negative, which means the effective field is parallel $+y$ instead of $\sigma \parallel -y$, and is likely dominated by $H_\mathrm{Oe}$. The SOT fields are quite different after including $X_u$ (compare the first and third data columns), especially $H_\mathrm{AD}$, which is reduced from 8.4 to approximately 3~mT. The effect of the $K_2$ anisotropy is relatively small (see the second and third columns). The exponent $p \approx 0.39$ is less than 1, similarly to Ref.~\citenum{Avci2018}. As for the role of the $X_{u0}$ amplitudes, let us take $A_{u0}$ in the third column of data (fifth row). Its 2HH component $A_{u0}K^{p}H^{-p}>0$ has opposite sign to the corresponding SOT contribution $R_P H_\mathrm{FL} H^{-1}$. Therefore the total $A$ plotted as a function of $H^{-1}$ would exhibit a shallow start to the net curve which becomes approximately linear later. The effect in $C$ is equivalent but with opposite sign as $C_{u0}<0$. This is indeed what we observed in Fig.~\ref{fig2}b,c).

   \begin{table}[!t]
 \label{tbl:1}
     \centering
     \begin{tabular}{ |p{3cm}|p{1.5cm}|p{1.5cm}|p{1.5cm}|  }
     \hline
         & 1: no $X_u$ & 2: no anis. & 3 \\
     \hline
        $\mu_0 K_2$ (mT) from $R_{xy}$  & 0.63(2)  & 0 & 0.63(2) \\
        $\mu_0 K_2$ (mT) from $R_{xx}$ & 1.45(3) & 0 & 1.45(3) \\
    \hline
        $\mu_0 H_\mathrm{FL}$ (mT)  & -0.37(1) & -0.46(1) & -0.49(2) \\
        $\mu_0 H_\mathrm{AD}$ (mT)  & 8.4(23) & 3.0(7) & 3.0(15) \\
    \hline
        $A_{u0}~(\mathrm{m\Omega})$  & 0 & 0.21(1) & 0.25(2)\\
        $B_{u0}~(\mathrm{m\Omega})$  & 0 & -0.16(4) & -0.18(8)\\
        $C_{u0}~(\mathrm{m\Omega})$  & 0 & -0.19(1) & -0.21(2)\\
        $D_{u0}~(\mathrm{m\Omega})$  & 0 & 0.31(5) & 0.32(9)\\
        $p$  & 0 & 0.43(3) & 0.39(6)\\
    \hline  
        Equations  & \ref{eq:R2wComps}, \ref{eq:R2w_X2b} & \ref{eq:R2wComps2}, \ref{eq:R2w_Xu0} & \ref{eq:R2wComps2}, \ref{eq:R2w_X2b}, \ref{eq:R2w_Xu0}\\
        Figures & \ref{Fig3} & - & \ref{Fig3} \\
    \hline
     \end{tabular}
     \caption{Relevant fit parameters without $X_u$ (1st data column), and with $H^{-p}$-type $X_u$, the latter without and with $K_2$ anisotropy (2nd and 3rd columns, respectively). All fits were based on $H$-dependence only. $H_\mathrm{FL}$ was shared as a parameter between the fitting functions to $A,A_2,C,C_2$. $p$ was shared by functions for $A,B,C,D$.}
 \end{table}

The symmetry argument discussed above would suggest that $|A_{u0}| = |C_{u0}|$ and $|B_{u0}| = |D_{u0}|$. However, since their origins may have different symmetries - consider that the AHE has no counterpart in $R_{xx}$ - we set no restrictions between their values. Looking at Table~\ref{tbl:1}, the values of $|A_{u0}|, |C_{u0}|$ are comparable, while $|D_{u0}|$ is almost twice as large as $|B_{u0}|$.

\textit{(2) The second approach: lengthwise modulation of the magnetization.} This approach is based on the fact that the transfer of angular momentum modifies the length of $\bm{M}$ \cite{Demidov2011,Kodama2024} in addition to its direction (via anti-damping torque). This concept has been expanded in recent publications, Refs.~\citenum{Noel2024a, Noel2024b}, where it is used to describe deviations of the 2HH data from expected, similarly to what we have presented above. Specifically, the length $M = |\bm{M}|$ is modified through current-induced magnon generation and annihilation in the form $-\Delta M (I) \bm{\sigma}\cdot \bm{M}$. Like in Refs.~\citenum{Avci2018, Xu2019}, electron-magnon scattering (MMR) plays a role in UMR. In this picture, part of $R_{xx}^{2\omega}$ is attributed to the current-modulation of MMR as $\Delta M$ oscillates with the current. However, AMR also contributes as its magnitude depends on $M$ \cite{Gerlach1930, Potter1931, Mott1936, Hamzic1978}, moreover, similar terms appear in $R_{xy}^{2\omega}$ due to PHE. A calculation of their contributions leads to the same $\varphi$-dependences in the 2$\omega$ signal as in Eq.~\ref{eq:R2w} (SI Sec.~\ref{subsec:UMR}), therefore they appear as the $A_u-D_u$ terms in Eqs.~\ref{eq:R2wComps2}. The major difference is that a predefined field-dependence is not imposed on $\Delta M$, rather, it is a free parameter at each $H$.

We employ this second approach to achieve a better description of our measurements and more accurate values for $H_\mathrm{FL,AD}$. The coefficients of the relative magnetization change $\Delta M $ are not arbitrary. As mentioned above, they originate from the magnetization-dependence of AMR, PHE and MMR (SI Sec.~\ref{subsec:UMR}). As a result, the UMR and related components of Eq.~\ref{eq:R2wComps2} are 
\begin{subequations}
\label{eq:R2w_Xu}
    \begin{eqnarray}
        A_u (H) &=&  R_P \Delta M \\
        B_u (H) &=& -R_P \Delta M  \\
        C_u (H) &=& -\frac{R_a}{2} \Delta M  \\
        D_u (H) &=& \frac{R_M}{2} \Delta M.
    \end{eqnarray}
\end{subequations}
$R_M$ describes the MMR contribution, however, it is difficult to extract from $R_0$ in Eq.~\ref{eq:Rb}. Therefore, we treat it as a free parameter independent of $H$. Besides $\Delta M$, $R_P$ and $R_a$ also depend on the field $H$ and change by 10\% in the studied range, as determined from fits to the first harmonic data (plotted in Fig.~\ref{SIfig:RP}).

\begin{figure}[!b]
\begin{center}
	\includegraphics[page=4]{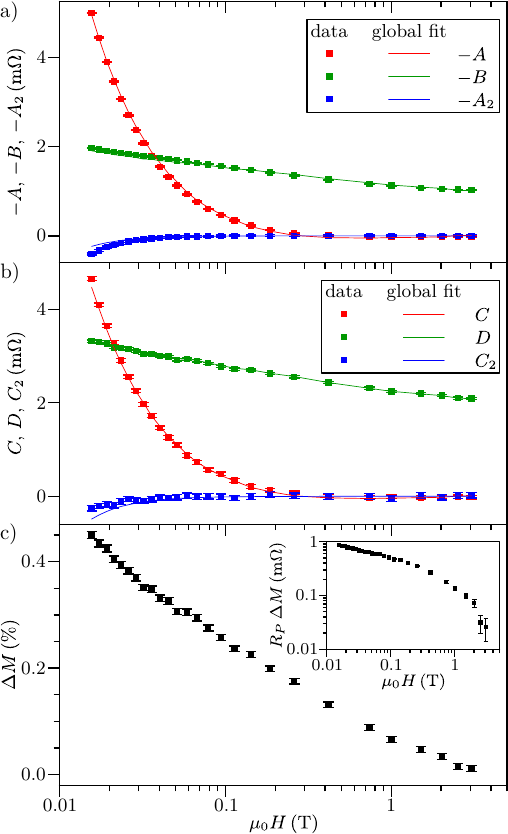}
	\caption{
    \textbf{Global fits with $\Delta M$-type corrections.} a) Amplitudes of $R_{xy}^{2\omega}$ and b) $R_{xx}^{2\omega}$ (symbols), following Eqs.~\ref{eq:R2w}, \ref{eq:R2w_X2}. Solid lines are the results of global fits (to $\varphi$ and $H$-dependence together) following Eqs.~\ref{eq:R2wComps2}, \ref{eq:R2w_X2b}, \ref{eq:R2w_Xu}. c) The relative magnetization change $\Delta M$ induced by the current decays towards zero at high fields. Inset:  $R_P \Delta M$ on a log-log scale.
	}\label{Fig4}\end{center}
\end{figure}

The curve fits using Eq.~\ref{eq:R2w_Xu} are shown in Fig.~\ref{Fig4}. In order to achieve a small relative error in the most sensitive parameter, $H_\mathrm{AD}$, we performed a global fit. This means the $\varphi$ and the $H$-dependence of the second harmonics $R_{xx,xy}^{2\omega}$ were optimized simultaneously instead of "local" fitting to $X(H)$. In general, the fitted curves are better than the first approach (Fig.~\ref{Fig3}), for example at high fields where the latter showed nonzero $A,C$. $\Delta M (H)$ is plotted in Fig.~\ref{Fig4}c): as expected, it decays fast to zero with increasing $H$. The log-log scale inset of $R_P \Delta M$ calculated from it illustrates that a power-law dependence of UMR-like corrections with $H$ may only work in the few-hundred-mT range. The fit to $D$ is also clearly better in the second approach. 

The results are summarized in the last column of Table~\ref{tbl:2}. $H_\mathrm{AD}\approx (0.49\pm 0.11)$~mT is significantly different from that estimated by the first approach ($(3.0 \pm 1.5)$~mT, see Table~\ref{tbl:1}, column three). The fact that the fits to $A,C$ according to the first approach do not converge to zero, together with the difference between the two results, demonstrates that the magnonic contribution cannot be described adequately by Eq.~\ref{eq:R2w_Xu0}, and that the second approach (Eq.~\ref{eq:R2w_Xu}) should be utilized to estimate SOT. The global fit greatly reduces the relative error of $H_\mathrm{AD}$ compared to the local fit ($(0.38\pm 0.96)$~mT, Table~\ref{tbl:2}, column one), but does not influence the result by a serious amount. Including the uniaxial anisotropy $K_2$ in the SOT analysis is crucial, as shown by comparing the third data column in Table~\ref{tbl:2} to the second. Without it, $H_\mathrm{AD}$ is almost doubled. Similarly, we found that not taking the field-dependence of $R_a$ and $R_P$ into account and treating them as constants would approximately double $H_\mathrm{AD}$. 

Let us compare the amplitudes $A_{u0}-D_{u0}$ in Table~\ref{tbl:1} to the coefficients of $\Delta M$ in Eq.~\ref{eq:R2w_Xu}, which is the most strongly field-dependent quantity there. We find that the signs match, moreover $C_{u0} \approx -A_{u0} $ similarly to $C_{u} \approx -A_{u} $ in Eq.~\ref{eq:R2w_Xu}. This is consistent with the symmetry argument since they come from the same effect (AMR/PHE). $B_u,D_u$ stand out since they are affected by AHE and MMR which lack a symmetrical counterpart in $R_{xx}$, $R_{xy}$. This supports the validity of the second approach. 

The question arises, can we select a field range where we can neglect UMR and its Hall counterpart and Eq.~\ref{eq:R2wComps} can be applied, as has generally been done? It is not the low-field regime, as $\Delta M$ is strongest here. Moreover, due to the fast decay of $A$ and $C$, a dense selection of $H$ points is required for accuracy. As for fitting $B$ only at high fields, the contribution of the AD-SOT and $B_u$ terms are both sub-$\mathrm{m\Omega}$ and remain relevant, as is also evident in the still observable field-dependence of $D$ which comes only from $\Delta M$. Consequently, a wide range of $H$ is required.

In order to compare our results with others in the literature, we analyzed the current-generated SOT fields. We have calculated the spin torque conductivity related to the AD torque, $\sigma_s = 2e \hbar^{-1}  \mu_0 H_\mathrm{AD} M_s t_\mathrm{Py} / \left( I_0 \overline{R}_{xx} / L \right)$ were $\mu_0$ is the vacuum permeability and $M_s = K \approx 5.45 \cdot 10^5~\mathrm{A/m}$ is the saturation magnetization (since crystalline anisotropy is negligible \cite{Yin2006}), which is comparable to the literature \cite{Ounadjela1988}. The denominator is the peak electric field calculated from the peak amplitude of the AC current $I_0 = \sqrt{2}\cdot 2.5~$mA, the average longitudinal resistance $\overline{R}_{xx}=R_0+R_a/2$ and the distance of side contacts $L$. In other words, this is the current density of the angular momentum absorbed in Py in the tilting of $\bm{M}$ by $\Delta \theta$, normalized by the electric field and converted to units of $\mathrm{S}$. The end result is $\sigma_s \approx (6.7 \pm 1.5) \cdot 10^4~$S/m, a value smaller but comparable to those in Ref.~\citenum{Liu2011b}. 

\begin{table}[!t]
\label{tbl:2}
    \centering
    \begin{tabular}{ |p{3cm}|p{1.5cm}|p{1.5cm}|p{1.5cm}|  }
    \hline
         & 1: local & 2: global, no anis. & 3: global \\
    \hline
        $\mu_0 K_2$ (mT) from $R_{xy}$  &  0.63(2) & 0 & 0.63(2) \\
        $\mu_0 K_2$ (mT) from $R_{xx}$  &  1.45(3) & 0 & 1.45(3) \\
    \hline        
        $\mu_0 H_\mathrm{FL}$ (mT) & -0.478(8)  & -0.453(1) & -0.471(1)  \\
        $\mu_0 H_\mathrm{AD}$ (mT) & 0.38(96) & 0.96(13) & 0.49(11) \\
    \hline
        Equations & \ref{eq:R2wComps2}, \ref{eq:R2w_X2b}, \ref{eq:R2w_Xu} & \ref{eq:R2wComps2}, \ref{eq:R2w_Xu} & \ref{eq:R2wComps2}, \ref{eq:R2w_X2b}, \ref{eq:R2w_Xu}\\
        Figures & - & - &  \ref{Fig4} \\
    \hline
    \end{tabular}
    \caption{Relevant fit parameters with the second approach, i.e. $\Delta M$-type $X_u$ corrections. The 1st column of data shows the result of a local fit, i.e. based on $H$-dependence only, similarly to Table~\ref{tbl:1}. In contrast, the 2nd and 3rd columns are results from global fits that take into account both $\varphi$ and $H$-dependence, without and with $K_2$ anisotropy, respectively.
    }
\end{table}

We also estimate the Oersted field $H_\mathrm{Oe} = I_\mathrm{Pt}/(2W) $. Since the current in Pt is the four-terminal voltage drop divided by the resistance of the Pt layer, $I_\mathrm{Pt} = I_0 \overline{R}_{xx} W t_\mathrm{Pt} / L \rho_\mathrm{Pt} \approx \sqrt{2}\cdot 1.75~$mA where we used $\rho_\mathrm{Pt} \approx 35~\mathrm{\mu \Omega cm}$ as the resistivity of a $t_\mathrm{Pt}=3$-nm-thick Pt film \cite{Dutta2017}, and as a result, we get $\mu_0 H_\mathrm{Oe}=0.389$~mT. Since this is comparable to $-\mu_0 H_\mathrm{FL}$ in Table~\ref{tbl:2}, the actual effective field of the FL torque is likely much smaller.

\section{Summary}

We have studied several Py/Pt bilayer Hall bars with in-plane rotations in a wide range of magnetic fields. We have found that in the second-harmonic Hall resistance the amplitudes $A,B$ contain unexpected field-dependent components. We attribute them to a transverse counterpart of the magnonic unidirectional magnetoresistance and show that they are critical in accurately estimating SOT fields. We demonstrated two approaches, \textit{(1)} one based on symmetry arguments, with a $H^{-p}$-type dependence already used to describe UMR in $R_{xx}^{2\omega}$. The other \textit{(2)} is based on the concept that the angular momentum transfer leading to the anti-damping torque and a perpendicular change in $\bm{M}$ also produces a lengthwise (parallel) change, $\Delta M$. This approach has been expanded in Refs.~\cite{Noel2024a, Noel2024b} and further developed here by including the shape anisotropy and the $H$-dependence of $R_{a}$ and $R_P$, which have a significant effect. The second approach yields a superior match to the observations as $\Delta M (H)$ is not described well by the phenomenological form $H^{-p}$. Fitting simultaneously to both $\varphi$ and $H$-dependence greatly reduced the relative error of $H_\mathrm{AD}$, the parameter that usually comes with a large error and appears the most sensitive to model corrections. 

Our comparison of the two approaches has shown that the second one provides the correct way to evaluate SOT. We expect that this method will lead to an improved description of second-harmonic magnetoresistance effects and a more accurate determination of spin-orbit torque fields. In turn, this will support the research of novel materials for the development of SOT-based devices.

\section{Methods}\label{sec:methods}

The Hall bars were fabricated of $6\,\mathrm{nm}$ permalloy ($\mathrm{Py}$, $\mathrm{Ni_{80}Fe_{20}}$) and $3\,\mathrm{nm}$ $\mathrm{Pt}$ via standard electron beam lithography and e-beam evaporation techniques. They were contacted by the deposition of $5\,\mathrm{nm}$ Ti and $50\,\mathrm{nm}$ Au in a subsequent step. The width of the Hall bars is $W = 4\,\mathrm{\mu m}$. The longitudinal separation of the side contacts is $L = 4\,\mathrm{\mu m}$ and their width is $w = 1\,\mathrm{\mu m}$ (see also Fig.~\ref{fig1}a)). For SOT analysis of Hall data, there is a geometrical correction factor \cite{Neumann2018} due to the current density distribution at the Hall voltage contacts. In our case $w/W \ll 1$, therefore, we expect a less than $10~\%$ correction which we neglect.

The electrical measurements were performed at room temperature with SR830, MFLI and HF2LI lock-in amplifiers and BASPI SP1004 low-noise/low-drift differential amplifiers. The current used for the resistance measurements in the main text was approximately $I_\text{rms} = 2.5\,\mathrm{mA}$ at $f = 132.43\,\mathrm{Hz}$. Based on measurements as a function of temperature on similar samples, we estimate that this induced at most $50\,\mathrm{K}$ temperature increase in the sample due to Joule heating. This is not expected to notably change the saturation magnetization \cite{Zhang2019, Devonport2018}. The used frequency $f$ is low enough for the second harmonic technique. Due to the high thermal diffusivity of Pt and Py, $\alpha >1.6\cdot 10^{-6}~\mathrm{m^2/s}$ \cite{Pt_web, Avery2015, Py_web}, the thermal diffusion length $l=\sqrt{\alpha/f}$ is on the order of hundreds of µm. This is much larger than the dimensions $L$ and $W$, moreover, heat generation is distributed over the device, therefore it can be considered in thermal equilibrium on the time scale of the measurement. This is confirmed by frequency-dependent measurements shown in the SI (Fig.~\ref{SIfig5}).

The samples were fixed on a rotating stage within a Janis 9TL-HRTB-30 superconducting magnet's bore. The actual field of the magnet was measured in-situ by a Hall sensor, or corrected by a calibration curve taken by the same sensor. In-plane rotations were performed in both directions at all fields so that electrical signal drift or mechanical failures such as jumps or torsion may be noticed. All rotation measurements were performed as a series of decreasing fields to ensure a single magnetic domain. Raw data of measurements in the main text are available online \cite{data}.

\bigskip

\section*{Acknowledgements}

This research was supported by the Ministry of Culture and Innovation and the National Research, Development and Innovation Office within the Quantum Information National Laboratory of Hungary (Grant No.\ 2022-2.1.1-NL-2022-00004 and UNKP-23-4-I-BME-36), by OTKA grants No.\ K138433 and K134437 and the NKKP STARTING grant No.\ 150232. We acknowledge funding from the Multi-Spin and 2DSOTECH FlagERA networks, the 2DSPIN-TECH Flagship project, the Alexander von Humboldt Foundation, the European Research Council ERC project Twistrain, and COST Action CA 21144 superQUMAP. M.H.D.G.\ acknowledges the funding from the European Union (ERC, 2D-OPTOSPIN, 101076932).

The authors thank J.\ G.\ Holstein, H.\ Adema, H.\ de Vries, A.\ Joshua, F.\ H.\ van der Velde, O.\ K.\ Christiansen, B.\ Horv\'ath, and M.\ Hajdu for their technical support. Hall bar fabrication was performed using Zernike NanoLabNL facilities (Groningen). We thank D.\ Szaller and S.\ Bord\'acs for their assistance with the electromagnet system.

\FloatBarrier 

\bibliography{references}

\balancecolsandclearpage


\onecolumngrid
\fontsize{11}{12}\selectfont
\appendix*

\counterwithin*{figure}{part}
\stepcounter{part}
\renewcommand{\thefigure}{S\arabic{figure}}

\counterwithin*{equation}{part}
\stepcounter{part}
\renewcommand{\theequation}{S\arabic{equation}}

\section*{Supplementary information}

\subsection{The Stoner-Wohlfarth model}\label{subsec:SW}

The Stoner-Wohlfarth (SW) model describes the classical energy formula of a static, uniform, single-domain ferromagnet with saturated magnetization. For a Py film with easy-plane anisotropy and rotational symmetry around the $z$ axis, we expect that the azimuthal angles of the magnetization and the field are equal ($\varphi=\varphi_H$, measured from the $x$ axis), while the polar angles ($\theta$, $\theta_H$, measured from the $z$ axis) may be different. The anisotropy energy in the simplest terms, in units of A/m, is
\begin{equation}\label{eq:SW1}
    \varepsilon \left( \theta, \varphi \right) = \frac{K}{2} \cos^2 \theta - \bm{H}\cdot \bm{M} = \frac{K}{2} \cos^2 \theta - H \left( \sin \theta_H \sin \theta \cos \left(\varphi_H-\varphi \right) + \cos \theta_H \cos \theta \right).
\end{equation}
Here $K$ is the anisotropy field which, in most FMs, contains both crystalline and demagnetization-related anisotropy, and is positive for an easy plane magnet. $H$ is the absolute value of the applied magnetic field. Once precession has been damped, the equilibrium direction of the magnetization is defined by $\partial_\theta \varepsilon =0$ and $\partial_\varphi \varepsilon=0$. If, as described in Eq.~\ref{eq:SW1}, there is no other anisotropy nor azimuthal SOT field component, $\partial_\varphi \varepsilon=0$ is solved by $\varphi = \varphi_H$ as expected, moreover, the second term is simplified to $- H \cos (\theta_H - \theta)$.

We can rephrase the condition of equilibrium as follows. The direction of the magnetization is defined by the balance of those components of the fields - including an effective anisotropy field - that are perpendicular to $\bm{M}$. Since $\varepsilon$ is in units of A/m, the sum of these components is equal to the negative derivative of the potential energy with respect to the angles, $-\partial_\theta \varepsilon $ and $-\partial_\varphi \varepsilon$. In equilibrium, these must be 0, which is consistent with the previous description.

In the presence of a weak spin-orbit torque field with polar and azimuthal components $H_{\theta, \varphi}$, the polar and azimuthal angles are modified to $\theta^\prime = \theta + \Delta \theta$ and $\varphi^\prime = \varphi + \Delta \varphi$ where we expect $\Delta \theta \propto H_\theta$ and $\Delta \varphi \propto H_\varphi$. Then the balance along the polar angle is given by setting the net perpendicular field to zero, i.e. the equilibrium condition is to be extended with the new field as $H_\theta - \partial_\theta \varepsilon \left( \theta^\prime, \varphi \right) = 0$. Assuming we are near equilibrium ($\partial_\theta \varepsilon \left( \theta, \varphi \right) = 0$) and also $\Delta \theta \ll \theta$,
\begin{equation}\label{eq:SW2}
    \Delta \theta \approx \frac{H_\theta}{H\cos \left( \theta_H-\theta \right) - K \cos 2\theta}.
\end{equation}
Similarly, the balance equation for the azimuthal angle near equilibrium ($\partial_\varphi \varepsilon \left( \theta, \varphi \right) = 0$ which leads to $\varphi = \varphi_H$ without SOT) is $H_\varphi - \partial_\varphi \varepsilon \left( \theta, \varphi^\prime \right) = 0$. This results in the relation
\begin{equation}\label{eq:SW3}
    \Delta \varphi \approx \frac{H_\varphi}{H \sin \theta_H}.
\end{equation}

\subsection{SOT via \texorpdfstring{$2 \omega$}{2ω} Hall measurements}\label{subsec:SOT}

In a high-symmetry system such as Pt, the polarization or current of a spin $\bm{\sigma}$ can generate both a field-like (FL) and an anti-damping-like (AD) torque in a magnet:
\begin{subequations}
	\label{eq:SOT} 
	\begin{eqnarray}
	\bm{\tau}_\mathrm{FL} &=& \gamma H_\mathrm{FL} \bm{M} \times \bm{\sigma}, 
	\\
	\bm{\tau}_\mathrm{AD} &=& \gamma H_\mathrm{AD} \bm{M} \times \left( \bm{M} \times \bm{\sigma} \right)
	\end{eqnarray}
\end{subequations}
where $H_\mathrm{FL,AD}$ are the magnitudes of the SOT fields proportional to the current, $\gamma$ is the gyromagnetic ratio, and $\bm{M}$ is the dimensionless magnetization vector. The effective fields are, therefore, $\bm{H}_\mathrm{FL} = H_\mathrm{FL} \bm{\sigma}$ and $\bm{H}_\mathrm{AD} = H_\mathrm{AD} \bm{M} \times \bm{\sigma}$. 

In Pt the spin comes from the spin Hall effect, therefore, $\bm{\sigma} \parallel -y$ at the Py/Pt interface. In an in-plane field rotation in Py, $\theta = \theta_H =\pi/2$ and $\bm{M} = \left( \cos \varphi , \sin \varphi , 0 \right)$. Then the polar SOT field appearing in Eq.~\ref{eq:SW2} is $H_\theta = H_\mathrm{AD} \cos \varphi$ which is parallel with the $z$ axis (and the torque $\tau_\mathrm{AD}$ is in the $xy$ plane), therefore it will have a measurable effect via the AHE. The azimuthal field in Eq.~\ref{eq:SW3} is $H_\varphi = -H_\mathrm{FL} \cos \varphi$ and is in the $xy$ plane, detectable via the PHE. In case of a low-frequency AC current these are multiplied by a time-dependence of the form $\sin \omega t$. Since $\sin^2 \omega t = (1- \cos 2\omega t ) / 2$, an out-of-phase component is generated in the 2$\omega$ voltage, from which we can calculate the SOT-related resistance. With the assumption $\Delta \theta, \Delta \varphi \ll 1$, it is of the general form
\begin{equation}
    \label{eq:SI:R2w0}
    R^{2\omega,\mathrm{SOT}} = -\frac{1}{2} \left( \frac{\partial R}{\partial \varphi} \Delta \varphi + \frac{\partial R}{\partial \theta} \Delta \theta \right)
\end{equation}
which applies to both $R_{xy}$ and $R_{xx}$. Following Eq.~\ref{eq:R} in the main text, and disregarding in-plane anisotropy,
\begin{subequations}
	\label{eq:SI:R2w} 
	\begin{eqnarray}
        R_{xy}^{2\omega,\mathrm{SOT}} &=& -R_P \frac{H_\varphi}{H} \cos 2\varphi + \frac{R_A}{2} \frac{H_\theta}{H+K} = R_P \frac{H_\mathrm{FL}}{H} \cos 2\varphi \cos \varphi + \frac{R_A}{2} \frac{H_\mathrm{AD}}{H+K} \cos \varphi ,
	\label{eq:SI:R2wa}
	\\
	R_{xx}^{2\omega,\mathrm{SOT}} &=& \frac{R_a}{2} \frac{H_\varphi}{H} \sin 2\varphi = \frac{-R_a}{2} \frac{H_\mathrm{FL}}{H} \sin 2\varphi \cos \varphi
    \label{eq:SI:R2wb}
	\end{eqnarray}
\end{subequations}
which is the SOT contribution to Eqs.~\ref{eq:R2w} and \ref{eq:R2wComps}. 

We may include additional terms in $R_{xx}(\theta,\varphi)$ such as the spin Hall or geometrical size effects. However, since $R_{xx}$ remains quadratic around $\theta = \pi/2$, its second harmonic is insensitive to the modulation $\Delta \theta$ to the first order.

\subsection{Additional experimental data}\label{subsec:extras}

\begin{figure*}[!tbh]
\begin{center}
	\includegraphics[page=5]{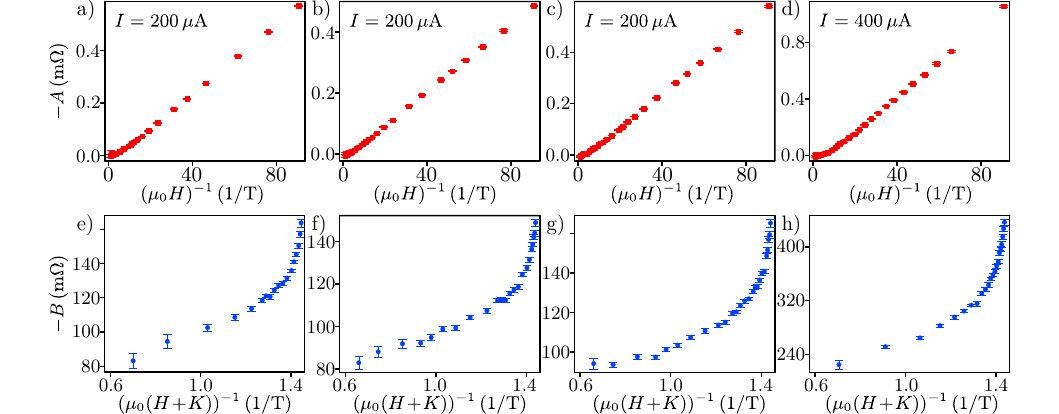}
	\caption{
	a-d) Coefficient $A$ as a function of $(\mu_0 H)^{-1}$ measured on four other samples.
	f-h) Coefficient $B$ as a function of $(\mu_0(H + K))^{-1}$ on the same set of samples, using $\mu_0 K = 0.68\,\mathrm{T}$. $K$ was similar between the samples with relatively little variation. Using the same value for simplicity in plotting doesn't affect the shape of the data and our conclusions. The current for most measurements was $200~\mathrm{\mu A}$ except in panels (d,h) where it was $400~\mathrm{\mu A}$.
	}\label{SIfig2}\end{center}
\end{figure*}

2HH data analysis on additional samples is shown in Fig.~\ref{SIfig2}. All results demonstrate that corrections are needed to the SOT model in Eq.~\ref{eq:R2wComps}. 

\subsection{Possible corrections}\label{subsec:possible}

Here we list the possible corrections that were considered but do not help explain the deviations from Eq.~\ref{eq:R2wComps}.

\begin{enumerate}[wide, labelwidth=!, labelindent=0pt]
    \item Eq.~\ref{eq:R2wComps} is based on the small-SOT-field limit, i.e. the assumption that the modulations $\Delta \theta , \Delta \varphi$ of the magnetization direction are small. Both the inadequate fits based on Eq.~\ref{eq:R2wComps} and the good fits based on the methods detailed in the main text generally produce $\mu_0 H_\mathrm{AD} < 9~\mathrm{mT} \ll \mu_0 (H+K)$ and $\mu_0 |H_\mathrm{FL} | < 0.5~\mathrm{m T} \ll H$. As a result, $\Delta \theta < $1° and $\Delta \varphi < $2° (see SI Section~\ref{subsec:SW}) so the assumption is valid and there is no need for higher order series expansion with $\Delta \theta, \Delta \varphi$. 
    \item FL and AD torques from other spin orientations ($\bm{\sigma} \parallel x,z$ axes) produce different angle-dependence compared to Eq.~\ref{eq:R2w}. They can be excluded as the equation describes all our samples well with small corrections. Rather, the deviations lie in the field-dependence of our data.
    \item The ordinary Nernst effect produces an electric field $\bm{E} \propto \bm{\nabla} T \times \bm{H}$ and is, therefore, linear in $H$. However, it only contributes to coefficients $B,D$ in Eq.~\ref{eq:R2w} while $A,C$ are unaffected. Its effect on the fit quality to the former is negligible, and the change in $H_\mathrm{AD}$ is relatively small. 
    \item In the presence of finite $\bm{\nabla} T$, the thermal analogues of AMR and PHE, i.e. anisotropic magnetothermopower and the planar Nernst effect, may produce electric fields $\bm{E} \parallel  \bm{M} (\bm{\nabla}T \cdot \bm{M})$ with $\cos 2\varphi$, $\sin 2 \varphi$ terms in $R_{xx,xy}^{2\omega }$ \cite{Ki1966, Pu2006, Jayathilaka2015, Avci2015, Reimer2017, Zink2022}. This introduces additional angle-dependence compared to Eq.~\ref{eq:R2w}. However, when allowing for such terms in the fit, they are negligible compared to the rest, moreover, they are expected to be independent of $H$. 
    \item Spins generated by $\bm{\nabla} T$, together with a spin-charge conversion, would produce an electric field with $2\omega$. For example, in the spin-dependent Seebeck effect the out-of-plane $\bm{\nabla} T$ generates a parallel spin current polarized $\parallel \bm{M}$, and the inverse SHE enables voltage detection, so $V_\mathrm{SSE} \propto \theta_\mathrm{SH} \bm{\nabla}T \times \bm{M}$ where $\theta_\mathrm{SH}$ is the spin Hall angle \cite{Uchida2010, Slachter2010, Bauer2012, Avci2014}. However, similarly to ANE, it would contribute a field-independent term to coefficient $B$. As a result of this, as well as UMR, we used non-equal "ANE" components in coefficients $B$ and $D$ in the fits. 
    \item We considered the idea whether the Stoner-Wohlfarth modeling of the dominant easy-plane anisotropy (Eq.~\ref{eq:SW1}), which leads to Eq.~\ref{eq:R2wCompsb}, requires corrections. We have measured $R_{xy}(\theta_H)$ in the vicinity of $\theta_H=\pi/2$ for a series of $H$, and found that the results for $R_A$ and $K$ are consistent with those evaluated based on $R_{xy}(H_z)$ (Fig.~\ref{fig1}), therefore the model holds in this regard. The details can be found in SI Sec.~\ref{subsec:ooprot}.
    \item Chip misalignment can include a small, constant difference between $\varphi$ and $\varphi_H$, but it is easily accounted for by an offset during fitting. It can also include an inclination, so that when the chip is rotated in the magnet at nominally $\theta_H = \pi/2$, a small, $\varphi_H$-dependent $H_z$ field appears. However, this only leads to a small correction to Eq.~\ref{eq:R}, and does not affect the 2$\omega$ formulae to the first order (see SI Sec.~\ref{subsec:misal}).
    \item The size of the magnetization likely depends on the field $H$, therefore the anomalous Nernst effect $R_\mathrm{ANE} \propto \bm{\nabla}T \times \bm{M}$ is not constant, as usually assumed. However, the saturation region of AHE in Fig.~\ref{fig1}c) is quite linear, and $M$ probably changes less than 1\%, and its effect on the analysis may be negligible. Moreover, it does not explain the field-dependence of $A,C$.
\end{enumerate}

\subsection{Uniaxial anisotropy}\label{subsec:SW2}

During a field sweep the magnetization reorientation occurs in the zero-field region, which we estimate to be approximately $10\,\mathrm{mT}$ wide based on the sharp dip in $R_{xx,xy}(H)$ around $H=0$ (see Fig.~\ref{fig1}c)). Therefore, we limited our measurements to higher fields. The first harmonics are, at first glance, well fitted by Eq.~\ref{eq:R} (see Fig.~\ref{fig1}d)), which is consistent with the uniformity of the magnetization.

\begin{figure*}[!t]
\begin{center}
	\includegraphics[page=6]{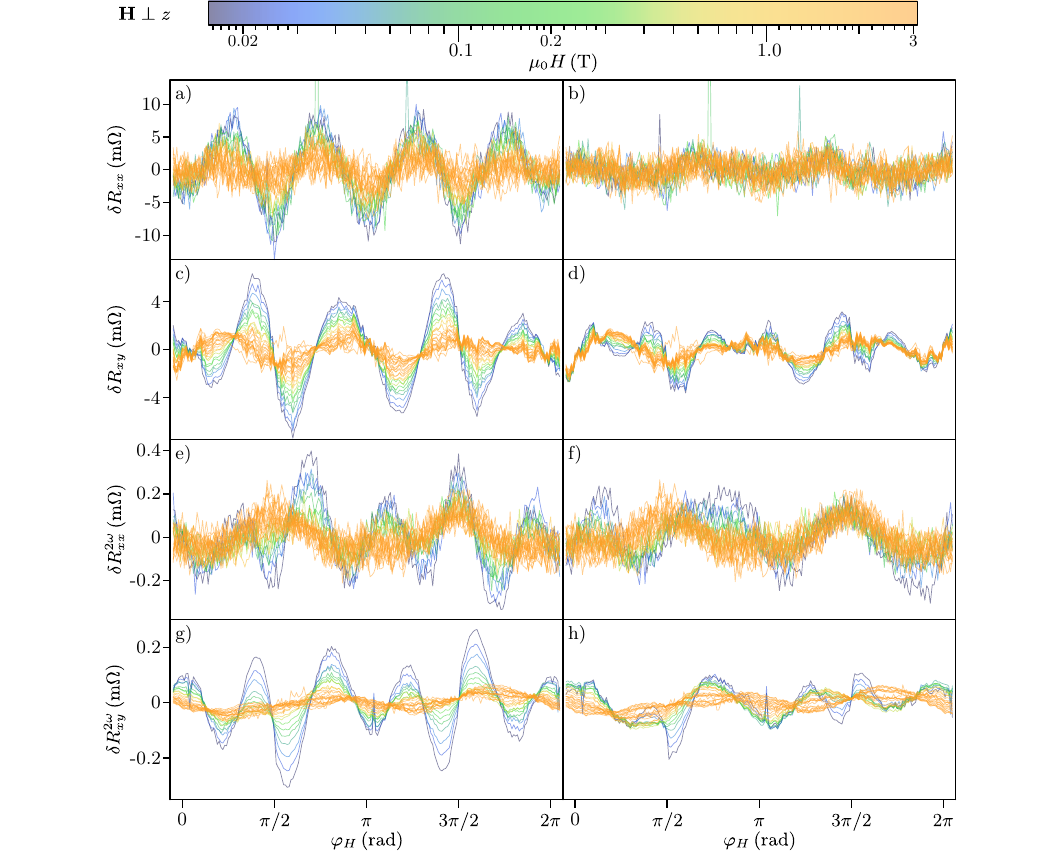}
	\caption{
    The residuals of fits to first and second-harmonics without (a,c,e,g) and with (b,d,f,h) $K_2$ uniaxial anisotropy.
	}\label{SIfig:res}\end{center}
\end{figure*}

However, as the residuals of these fits show in Fig.~\ref{SIfig:res}a,c), there is a systematic, $\pi/2$-periodic deviation that decays with the field. We attribute this to an in-plane uniaxial anisotropy.

An in-plane easy axis from shape anisotropy can realistically be present along the long, $I \parallel x$ axis of the Hall bar. Therefore we extend the SW model (Eq.~\ref{eq:SW1}) with the following term:
\begin{equation}\label{eq:SW4}
    \varepsilon_2 \left( \theta, \varphi \right) = -\frac{K_2}{2} \sin^2 \theta \cos^2 \left( \varphi - \varphi_2 \right)
\end{equation}
where $K_2>0$ is the characteristic anisotropy field strength and $\varphi_2$ is the easy-axis direction in the $xy$ plane relative to $x$. Assuming $K_2 \ll H$, there is a small deviation relative to the field angle $\varphi_H$. At $\theta=\pi/2$ it is
\begin{equation}\label{eq:SW5}
    \varphi \approx \varphi_H - \frac{K_2}{2H} \sin 2(\varphi_H - \varphi_2)
\end{equation}
rather than $\varphi = \varphi_H$, which is consistent with Ref.~\citenum{Macneill2017a}. The fits to the first harmonic data with Eq.~\ref{eq:R} using the above modification are significantly improved, as demonstrated by the residuals plotted in Fig.~\ref{SIfig:res}b,d). We have estimated that the characteristic anisotropy strength is around $\mu_0 K_2 \sim 0.6$ and 1.4~mT for $R_{xy}$, $R_{xx}$, respectively (see the first two rows in Table~\ref{tbl:1}). We attribute their variation to the difference in the local geometry at the voltage probes, i.e. a cross and a rectangle, and we use both values when fitting the second harmonics. The $\varphi_H$-dependent correction is as high as $K_2 / 2H \sim$3° at 16~mT. $\varphi_2$ is the easy axis direction, usually no more than 3° and approximated as $0$, i.e. the easy axis is indeed along the length of the Hall bar. Essentially, equations~\ref{eq:R},~\ref{eq:R2w} remain in effect but with a $\varphi(\varphi_H,H)$ function. 

Due to the extra term, Eq.~\ref{eq:SW2} evaluated at $\theta=\theta_H=\pi/2$ is modified to
\begin{equation}\label{eq:SW2b}
    \Delta \theta \approx \frac{H_\theta}{H + K + K_2 \cos^2 (\varphi - \varphi_2 )} \approx \frac{H_\theta}{H + K},
\end{equation}
essentially unchanged since $K_2$ is negligible compared to $K$. Also, Eq.~\ref{eq:SW3} changes to
\begin{equation}\label{eq:SW3b}
    \Delta \varphi \approx \frac{H_\varphi}{H + K_2 \cos 2 (\varphi - \varphi_2 )} \approx \frac{H_\varphi}{H} \left( 1 - \frac{K_2}{H} \cos 2 (\varphi_H - \varphi_2 ) \right)
\end{equation}
to first order since $K_2 \ll H$, which makes $K_2$ relevant at low $H$ in SOT, besides the fact that the fit abscissa of Eq.~\ref{eq:R2w} is modified by Eq.~\ref{eq:SW5}. 

Then the SOT components extended with anisotropy $K_2$ are:
\begin{subequations}
	\label{eq:SI:R2w2} 
	\begin{eqnarray}
        R_{xy}^{2\omega,\mathrm{SOT}} &=& R_P \frac{H_\mathrm{FL}}{H} \left( 1 - \frac{K_2}{H} \cos 2 (\varphi - \varphi_2 ) \right) \cos 2\varphi \cos \varphi   + \frac{R_A}{2} \frac{H_\mathrm{AD}}{H+K} \cos \varphi ,
	\label{eq:SI:R2w2a}
	\\
	R_{xx}^{2\omega,\mathrm{SOT}} &=& \frac{-R_a}{2} \frac{H_\mathrm{FL}}{H} \left( 1 - \frac{K_2}{H} \cos 2 (\varphi - \varphi_2 ) \right) \sin 2\varphi \cos \varphi 
    \label{eq:SI:R2w2b}
	\end{eqnarray}
\end{subequations}
where $\varphi$ must follow Eq.~\ref{eq:SW5}. As a result, Eqs.~\ref{eq:R2wa},~\ref{eq:R2wb} are to be extended with new terms:
\begin{subequations}
    \label{eq:R2w_X2}
        \begin{eqnarray} 
            A_2 \cos 2 \varphi \cos \varphi \cos 2 (\varphi - \varphi_2 ) \label{eq:R2w_A2}
            \\
            C_2 \sin 2 \varphi \cos \varphi \cos 2 (\varphi - \varphi_2 ) \label{eq:R2w_C2}
        \end{eqnarray}
\end{subequations}
for $R_{xy}$ and $R_{xx}$, respectively. The coefficients are
\begin{subequations}
     \label{eq:R2w_X2b}
         \begin{eqnarray}
             A_2 &=& - R_P \frac{H_\mathrm{FL} }{H} \cdot \frac{K_2}{H}, \label{eq:R2w_A2b}
             \\
             C_2 &=& \frac{R_a}{2} \frac{H_\mathrm{FL} }{H} \cdot \frac{K_2}{H}. \label{eq:R2w_C2b}
         \end{eqnarray}
\end{subequations}
The reestimated amplitudes $A$--$D$ are shown in Fig.~\ref{fig2}b-e), along with $A_2, C_2$. While allowing for this effect does not explain the major issue in the field-dependence of the second harmonics, it essentially eliminates the deviation in $\varphi_H$-dependence at low fields. This is best shown in Fig.~\ref{fig2}a) at 16~mT: the fits with finite anisotropy (solid lines) nicely follow the measured curves compared to those without anisotropy (dashed lines). The improvement in the fit residuals in Fig.~\ref{SIfig:res}f,h) is also evident, and indeed decays fast with increasing $H$. 

We have considered additional corrections to the angle-dependence of the first and second harmonics. For instance, a small, $\mathrm{m\Omega}$-scale, $2\pi$-periodic signal can be extracted from the first harmonic measurements which we attribute to a slight misalignment of the chip relative to the magnet. The $\pi/3$-periodic signal that can be best distinguished in $\delta R_{xy}$ ($K_2 \neq 0$) in Fig.~\ref{SIfig:res}d) at low $H$ is likely from an anisotropy with fourfold symmetry, which is realistic considering the cross shape of the device at the Hall contacts. It is modeled using a term $ -\frac{K_4}{4} \cos^2 2 ( \varphi - \varphi_4 )$ in the SW energy functional, and it clearly improves the first harmonic fits (not shown). However, $K_4$ is $3-5$ times weaker than $K_2$ (while $\varphi_4 \approx 0$) and does not visibly improve the $2\omega$ fits. A $2\pi/3$-periodic component, best visible in $\delta R_{xx}$, $\delta R_{xy}$ in Fig.~\ref{SIfig:res}b,d) at high $H$ since the correction from $K_2$ decays as $1/H^2$, may be due to to high-order components in AMR, PHE \cite{Ritzinger2023}. 

Regarding $\delta R_{xx}^{2\omega}$, $R_{xy}^{2\omega}$ ($K_2 \neq 0$) in Fig.~\ref{SIfig:res}f,h), $\pi$-periodic components are still present which we attribute to the anisotropic magnetothermopower and the planar Nernst effect. These corrections barely affect $H_\mathrm{AD}$ and $H_\mathrm{FL}$. 

These analyses were performed based on the second approach in the main text (UMR as a modulation to the magnetization), which produces better fits and is detailed in SI. Sec.~\ref{subsec:UMR}.

\subsection{SF-UMR as a modulation of the magnetization}\label{subsec:UMR}

Magnon generation and annihilation depend on the angle of the magnetization and the current-induced spins via $\bm{\sigma} \cdot \bm{M} = -\sin \theta \sin \varphi$ for the case of Py/Pt interface where $\bm{M}$ is the dimensionless magnetization. Therefore, at $\theta = \pi/2$, following the formalism of Refs.~\citenum{Noel2024a, Noel2024b}, the magnetization is

\begin{equation}
\label{eq:dM}
    M(I) = 1 - \Delta M (I) \bm{\sigma}\cdot \bm{M} =  1 + \Delta M(I) \sin \varphi.
\end{equation}
$\Delta M(I) \geq 0$ measures the relative decrease when $\bm{\sigma} \parallel \bm{M}$ and magnons are generated, and is an odd function of the current $I$, assumed to be primarily linear for low currents. Since we are at room temperature, $M$ can be greater than 1 with an applied current in case of magnon annihilation ($\bm{\sigma} \parallel -\bm{M}$). 

The amplitudes of AMR ($R_a$) and PHE ($R_P$) are approximately proportional to $M^2$ \cite{Gerlach1930, Potter1931, Mott1936,Hamzic1978}, therefore we can extend their expressions in Eq.~\ref{eq:R} as

\begin{subequations}    
    \begin{eqnarray}
        R_\mathrm{PHE} &=& R_P M^2 \sin 2\varphi \approx R_P (\sin 2\varphi + 2 \Delta M \sin 2\varphi \sin \varphi)
        \label{eq:dMa}
        \\
        R_\mathrm{AMR} &=& R_a M^2 \cos^2 \varphi \approx R_a ( \cos^2 \varphi + 2 \Delta M \cos^2 \varphi \sin \varphi)  
        \label{eq:dMb}
    \end{eqnarray}
\end{subequations}    

\begin{figure*}[!tbh]
\begin{center}
	\includegraphics[page=7]{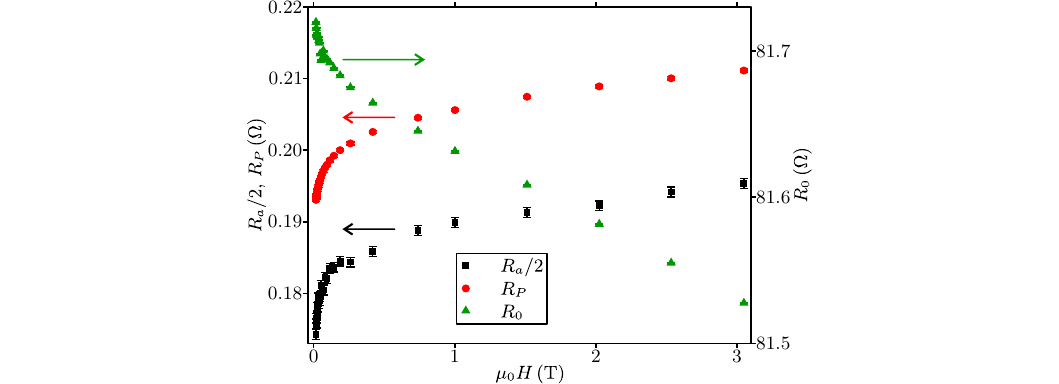}
	\caption{
    The field-dependence of AMR ($R_a/2$) and PHE ($R_P$) in the first harmonic resistances. The $\varphi_H$-independent background $R_0$ in $R_{xx}$ according to Eq.~\ref{eq:R} demonstrates the MMR.
	}\label{SIfig:RP}\end{center}
\end{figure*}

The magnon magnetoresistance $R_\mathrm{MMR}$ \cite{Mihai2008} from isotropic electron-magnon scattering is also relevant. It contributes to the approximately linear field-dependence of $R_0$ (see Eq.~\ref{eq:Rb}) which is plotted in Fig.~\ref{SIfig:RP}. We expect that it is also affected by current-induced or -annihilated magnons, therefore, we describe its correction as $ R_M \Delta M \bm{\sigma}\cdot \bm{M} = -R_M \Delta M \sin \varphi$ where $R_M > 0$. The amplitude of AHE ($R_A$) can also be modified by $\Delta M$ but it does not come into play since it is multiplied by $\cos \theta = 0$ according to Eq.~\ref{eq:Ra}. 

As a result, since $\Delta M \propto I$, all of the above contribute to the 2$\omega$ signal as 
\begin{subequations}
    \label{eq:SI:dM}
    \begin{align}
        R_{xy}^{2\omega,\mathrm{UMR}} &= -R_P \Delta M \sin 2 \varphi \sin \varphi = -R_P \Delta M ( - \cos 2 \varphi \cos \varphi + \cos \varphi ) \\
        R_{xx}^{2\omega,\mathrm{UMR}} &= - R_a \Delta M \cos^2 \varphi \sin \varphi + \frac{R_M}{2} \Delta M \sin \varphi = -\frac{R_a}{2} \Delta M  \sin 2\varphi \cos \varphi + \frac{R_M}{2} \Delta M \sin \varphi 
    \end{align}
\end{subequations}
where we included the $-1/2$ factor discussed at Eq.~\ref{eq:SI:R2w0}. The field-dependence is dominated by $\Delta M$, although $R_P, R_a$ also depend on $H$. As shown in Fig.~\ref{SIfig:RP}, the latter show a monotonous increase.

$\Delta \theta$ and $\Delta M$ represent the relative perpendicular and parallel modulation of $\bm{M}$ due to angular momentum transfer. At the lowest $H$ they are approximately 0.0007~rad and 0.004, respectively, i.e. the value of $\Delta M$ is an order of magnitude stronger. This highlights the importance of $\Delta M$ in SOT analysis and indicates that further theoretical work may be necessary for an accurate description of spin transfer. However, this would require careful consideration of the current-induced (non-equilibrium) magnon populations and energies, which is beyond the scope of this manuscript.

We note that, since the Hall and longitudinal measurements probe different areas of the Hall bar (between transverse Hall voltage probes and between side contacts), the deviation from $R_a = 2 R_P$ estimated from $R_{xy}$ and $R_{xx}$ suggests that the sample may be slightly inhomogeneous. Therefore, in order to improve the 2$\omega$ fits, we have tried allowing for different $H_\mathrm{FL}$ parameters in fitting $A,A_2$ and $C,C_2$. While the fits are slightly better compared to Fig.~\ref{Fig4} (not shown), the change in fit parameters is negligible.

\subsection{The effect of chip misalignment for in-plane rotations}\label{subsec:misal}

\begin{figure*}[!tbh]
\begin{center}
	\includegraphics[page=8]{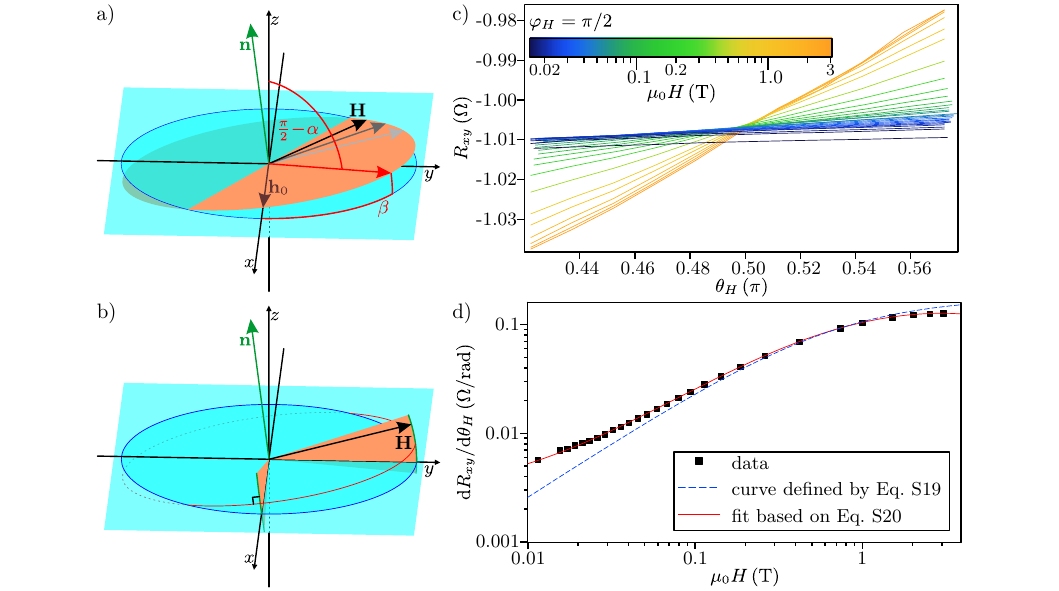}
	\caption{
	a) Illustration of chip misalignment during in-plane field rotation with $\varphi_H$ ($\theta_H \approx \pi/2$), in the frame of the chip. The chip ($xy$) plane is light blue. The path of the field vector follows the orange disc. The red arrow is the $\bm{H}$ field direction at maximal $H_z$. The green vector is the circle's normal vector $\bm{n}$. b) Illustration of out-of-plane field rotations with $\theta_H$ in the presence of the same misalignment when $\theta_H$ is near $\pi/2$. The orange sectors represent the path of the $\bm{H}$ vector at $\varphi_H = 0$ and $\pi/2$. c) Out-of-plane rotation measurements with $\theta_H$ near $\pi/2$ at a series of fields at $\varphi_H = \pi/2$. d) The slope of the measurements in c) (black symbols), the expectation based on the $H_z$-sweep shown in Fig.~\ref{fig1}c) (blue dashed line, see Eq.~\ref{eq:ooprot1}) and a fit following Eq.~\ref{eq:ooprot2} (red line).
    }\label{SIfig1}\end{center}
\end{figure*}

In general, the chip plane ($xy$) is likely to be slightly misaligned to (not perfectly parallel with) the electromagnet axis (the applied field) during in-plane rotation with $\varphi_H$, and $\theta_H = \pi/2$ is not exact. When the chip is rotated in its $xy$ plane relative to the field, this means that the field $\bm{H}$ relative to the chip "wobbles" with a $2\pi$ period. The path of the field vector is illustrated in Fig.~\ref{SIfig1}a) by the orange disc. Below we calculate the actual field angle $\theta_H$ as a function of the rotation variable $\varphi_H$, and how it affects the equilibrium $\theta$ and our previous equations.

Let us assume that, relative to the chip coordinate system, the highest $H_z$ component is reached when $(\theta_H, \varphi_H) = (\pi/2 - \alpha, \beta)$ where $0 < \alpha \ll 1$, i.e. misalignment is small, as shown in Fig.~\ref{SIfig1}a). Then the normal vector $\bm{n}$ of the orange disc (green arrow) has angles $(\alpha, \beta + \pi)$, therefore $\bm{n} = \left( -\sin \alpha \cos \beta, -\sin \alpha \sin \beta, \cos \alpha \right)$. When the field is mostly parallel with $x$, i.e. $H_y = 0$ and $\varphi_H = 0$, we define its direction as a $\bm{h_0}$ unit vector. From $\bm{n} \cdot \bm{h_0}=0$ we get $\bm{h_0} \approx \left( 1, 0, \cos \beta \tan \alpha \right)$.

In order to calculate the field direction $\bm{h}(\varphi_H)$ during rotation, we define $\bm{w} = \bm{n} \times \bm{h_0}$ which is the tangent of the orange disc at $\varphi_H=0$, therefore $\bm{h}(\varphi_H) = \bm{h_0} \cos \varphi_H + \bm{w} \sin \varphi_H$. Since $\alpha \ll 1$, $ \bm{w} \approx \left( 0, 1, \alpha \sin \beta \right)$ and $\bm{h} \approx \left( \cos \varphi_H, \sin \varphi_H, \alpha \cos (\beta - \varphi_H ) \right)$. By defining the function $d(\varphi_H) = \alpha \cos (\beta - \varphi_H )$, we can write the field's polar angle as approximately $\theta_H(\varphi_H) \approx \pi/2 - d(\varphi_H)$. 

Based on the SW model described above, we have calculated the corrected magnetization angles in equilibrium. While $\varphi$ is unaffected, the polar angle is
\begin{equation}
    \theta(\varphi_H,H) \approx \pi/2 - d(\varphi_H) \frac{H}{H + K + K_2 \cos^2 (\varphi - \varphi_2 )} \approx  \pi/2 - d(\varphi_H) \frac{H}{H + K}.
    \label{eq:theta}
\end{equation}
We calculate that the SOT modulations $\Delta \theta, \Delta \varphi$ do not depend on $d(\varphi_H)$ to the first order. Therefore, the only effect is a small correction to $R_{xy}$ due to the AHE. Using the series expansion of $\cos \theta$, it is $R_A d(\varphi_H) \frac{H}{H + K}$. It is $2\pi$-periodic after neglecting the correction from $K_2 \ll K$. As discussed in relation to Fig.~\ref{SIfig:res}, a $\mathrm{m \Omega}$-scale component with $2\pi$ period can indeed be extracted from $R_{xy}$. Besides AHE, the ordinary Hall effect $R_\mathrm{O}H \cos \theta_H$ also provides a correction term. Since $R_\mathrm{O} \approx 7~\mathrm{m\Omega/T}$ is already small, its contribution ($d(\varphi_H) R_\mathrm{O} H$) is negligible. 

\subsection{Out-of-plane rotations}\label{subsec:ooprot}

As we have discussed before in detail, the first harmonic resistances for in-plane rotations have been measured at a series of fields and, using Eq.~\ref{eq:R} and the SW-model, we have confirmed the easy-plane nature of the magnetization in Py, albeit with a weak uniaxial anisotropy. Below we confirm the predictions of the SW model (Eq.~\ref{eq:SW1}) regarding the effect of out-of-plane (polar, $H_\theta$) field components, which is relevant for the exact form of the term with $H_\mathrm{AD}$ in Eq.~\ref{eq:R2wCompsb}. 

We have performed Hall measurements while slightly tilting the chip, i.e. changing $\theta_H$ in the vicinity of $\theta_H=\pi/2$ at a nominally constant $\varphi_H$. The path of the field vector relative to the chip is illustrated in Fig.~\ref{SIfig1}b) by orange sectors, while the measurements are plotted in panel c). The SW model predicts that around $\theta_H=\pi/2$ the polar angle of the magnetization is
\begin{equation}
    \theta = \pi/2 + \frac{H}{H+K} (\theta_H - \pi/2).
\end{equation}
Misalignment can offset $\theta_H$ by a small $d(\varphi_H)$ as calculated in the previous section. However, since $d(\varphi_H)$ is small and independent of $\theta_H$, it does not affect the derivative of $R_{xy}(\theta_H)$ which is the quantity we are interested in. Thus, using Eq.~\ref{eq:R}, we expect the slope of the Hall data in these rotations to be
\begin{equation}
    \label{eq:ooprot1}
    dR_{xy}/d\theta_H = - R_A \frac{H}{H+K} >0.
\end{equation}
PHE ($\propto \sin^2 \theta$) does not contribute in the studied range of $\theta_H$ as it has an extremum here. We show the slopes of the linear fits to measurements in Fig.~\ref{SIfig1}d) in black symbols. We also plot the curve predicted by Eq.~\ref{eq:ooprot1} as a dashed blue line, using $R_A \approx - 179\,\mathrm{m \Omega}$ and $\mu_0 K \approx 0.685\,\mathrm{T}$ (from the $H_z$-dependence in Fig.~\ref{fig1}c)), which clearly deviates from the data. 

We have found that the following equation provides a better fit (red line):
\begin{equation}
\label{eq:ooprot2}
    dR_{xy}/d\theta_H = -R_A \frac{H}{H+K} - b\cdot \mu_0 H + c.
\end{equation}
The fit results are the following: $R_A \approx - 189(3)\,\mathrm{m \Omega}$, $\mu_0 K \approx 0.696(10)\,\mathrm{T}$, $b \approx 9.7(7)\,\mathrm{m\Omega / T}$, and $c\approx 2.7(1) \,\mathrm{m \Omega}$. We attribute coefficient $b$ to the ordinary Hall effect which has a comparable magnitude (see Fig.~\ref{fig1}c)), and the constant offset $c$ to the chip misalignment $\alpha$. How can misalignment still affect the results? Fig.~\ref{SIfig1}b) illustrates the movement of the $\bm{H}$ vector by orange sectors in this configuration, showing that in the presence of misalignment ($\alpha \neq 0$) $\bm{H}$ can have in-plane components. This small change $\delta \varphi_H$ of the azimuthal angle as a function of $\theta_H$ introduces a small term to $R_{xy}(\theta_H)$ through the PHE which is independent of $H$, producing a constant term in $dR_{xy}/d\theta_H$. 

The results for $R_A$ and $K$ are consistent with the values estimated in Fig.~\ref{fig1}c), therefore, we conclude that the SW model does not require corrections in this respect.

\subsection{Frequency-dependence}\label{subsec:freqdep}

\begin{figure*}[!tbh]
\begin{center}
	\includegraphics[page=9]{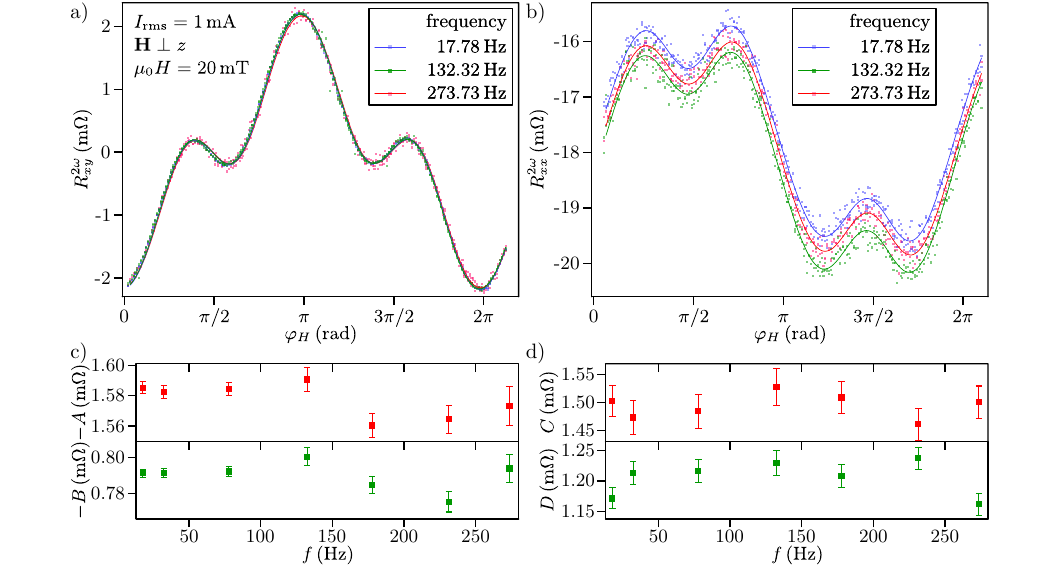}
	\caption{The frequency-dependence of the signal at $\mu_0 H = 20\,\mathrm{mT}$ using $1\,\mathrm{mA}$ rms current. a) $R_{xy}^{2\omega}$ and b) $R_{xx}^{2\omega}$ with in-plane field rotation ($\theta = \theta_H = \pi/2$) for three selected frequencies. The solid lines are fits according to Eqs.~\ref{eq:R2w} and \ref{eq:R2w_X2}. c) The frequency-dependence of the second-harmonic amplitudes $-A,-B$ and d) $C,D$. 
    }\label{SIfig5}
\end{center}
\end{figure*}

In order to confirm that our analysis is not distorted by the used frequency, we have checked the frequency dependence of the signal for several values between $10-300\,\mathrm{Hz}$. The results are summarized in Fig.~\ref{SIfig5}.

We performed the second harmonic resistance measurements as a function of $\varphi$ at $I_\text{rms} = 1 \,\mathrm{mA}$ and an in-plane field $\mu_0 H = 20\,\mathrm{mT}$. We fitted the data according to Eqs.~\ref{eq:R2w} and \ref{eq:R2w_X2} (Fig.~\ref{SIfig5} a-b)), taking the in-plane anisotropy into account. The extracted coefficients $A,B,C,D$ are plotted as a function of frequency in Fig.~\ref{SIfig5} c-d). They show a negligible frequency dependence, indicating the device is in thermal equilibrium on the time scale of the measurement, likely due to its small size and the high thermal diffusivity of Py and Pt.

\subsection{Alternative data-visualization}\label{subsec:refplots}

\begin{figure*}[!tbh]
\begin{center}
	\includegraphics[page=10]{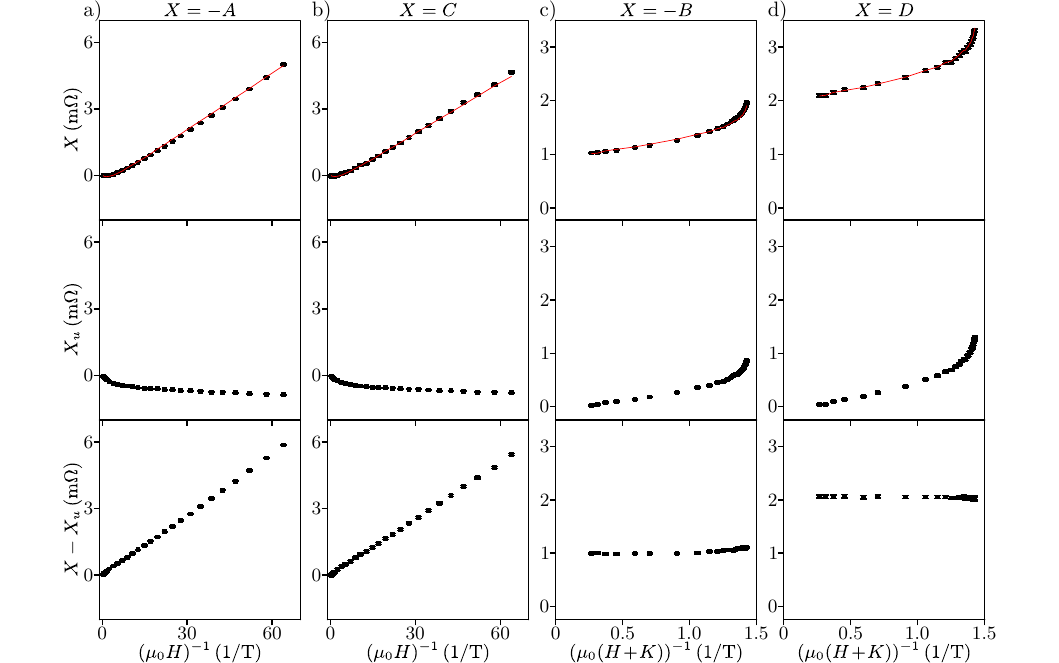}
	\caption{First row: the amplitudes of $R_{xy}^{2\omega}$ and $R_{xx}^{2\omega}$ based on Eqs.~\ref{eq:R2w}, \ref{eq:R2w_X2} are plotted as a function $(\mu_0 H)^{-1}$ and $(\mu_0 (H + K))^{-1}$. The fits following $\Delta M$-type UMR (Eqs.~\ref{eq:R2wComps2}, \ref{eq:R2w_Xu}, seen on log-scale in Fig.~\ref{Fig4}) are shown in red solid lines. In the second row are the separate UMR components based on the fits. In the third row is the difference between the points in the first two rows. 
    }\label{SIfig6}
\end{center}
\end{figure*}

We plot the amplitudes $-A,C,-B,D$ and their fits (solid lines) in Fig.~\ref{SIfig6} as a function of the inverse field, similarly to Fig.~\ref{fig2}b-e). The UMR-related corrections based on the second approach (Eq.~\ref{eq:R2w_Xu}) are also plotted (second row), as well as their difference (third row) which should follow Eq.~\ref{eq:R2wComps}. 

\subsection{Current-dependence of effective fields}\label{subsec:currdependence}

\begin{figure*}[!tbh]
\begin{center}
	\includegraphics[page=11]{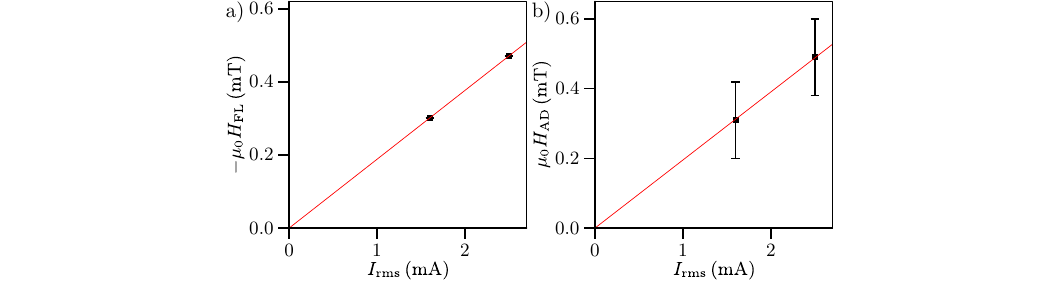}
	\caption{The field-like (a) and damping-like torques (b) as a function of rms current. Red lines are linear fits that intercept the origin.
    }\label{SIfig7}
\end{center}
\end{figure*}

We have repeated the set of measurements at another current, 1.6~mA. We plot the SOT fields extracted via the second approach with a global fit in Fig.~\ref{SIfig7}, and include the results with 2.5 mA from the main text (final column of Table~\ref{tbl:2}). Both $H_\mathrm{FL}$ and $H_\mathrm{AD}$ are approximately linear with current, according to expectations.

\end{document}